\documentclass[useAMS,usenatbib]{mn2e}
\usepackage{times}
\usepackage{amssymb}

\newcommand{\kms}{km~s$^{-1}$}
\usepackage{graphics,epsfig}

\def\kl{{\rm k} \lambda}
\def\mnras{MNRAS}

\title[HI Power spectrum of DDO 210]{Power spectrum of HI intensity
 fluctuations in DDO 210}
\author[
Begum, Chengalur \& Bharadwaj
]
{
Ayesha Begum$^{1}$\thanks{E-mail:ayesha@ncra.tifr.res.in},
Jayaram N. Chengalur$^{2}$ and
Somnath Bharadwaj$^{3}$\thanks{E-mail:somnathb@iitkgp.ac.in}
\\
\\
$^{1}$National Centre for Radio Astrophysics, Post Bag 3, Ganeshkhind, Pune 411 007, India\\
$^{2}$ ATNF/CSIRO P. O. Box 76, Epping NSW 1710, Australia. On leave from
       NCRA/TIFR\\
$^{3}$ Department of Physics and Meteorology \&
 Centre for Theoretical Studies ,  IIT Kharagpur, Pin: 721 302 ,
  India. \\
}
\begin{document}

\date{}


\maketitle

\label{firstpage}

\begin{abstract}
We measure the power spectrum of HI intensity fluctuations in
the extremely faint (M$_{\rm B}~\sim~-10.9$) dwarf galaxy 
DDO~210 using a visibility based estimator that is well suited 
in very low signal to noise ratio regimes.  DDO~210's HI power 
spectrum is well fit by  a power law ${\rm P}_{\rm HI}(U)=
A U^{\alpha}$ with  $\alpha=-2.75 \pm0.45$  over the length-scales 
$80 \, {\rm pc}$ to $500 \, {\rm pc}$.  We also find that
the power spectrum does  not change with an increase in the 
velocity channel width,  indicating that the measured
fluctuations correspond mainly to density fluctuations.
However, Kolmogorov turbulence (with a velocity structure 
function spectral slope of 2/3) cannot be ruled out from 
the present observations. The value of the slope $\alpha$ is 
similar to that obtained in the Milkyway. In contrast to the Milkyway, 
DDO~210 has three orders of magnitude less HI, no spiral arms,
and also no measurable ongoing star formation.  The fact that 
the power spectrum slope is nonetheless similar in these 
two galaxies (and also similar to the values measured for
the LMC and SMC) suggests that there is some universal,
star formation independent, phenomenon responsible for 
producing fine scale structure in the gas.
\end{abstract}


\section{Introduction}
\label{sec:intro} 

      Evidence has been mounting in recent years that turbulence plays 
an important role in determining the physical conditions of the 
neutral ISM as well as for generating the  hierarchy of structures 
seen in it ( see Elmegreen \& Scalo (2004) and Scalo \& Elmegreen
(2004) for recent reviews). Observational evidence includes the 
fact that the fluctuation 
power spectrum of a variety of tracers (HI 21cm emission intensity, 
HI 21cm optical depth, dust emission) is a scale free power law.  
The slope of the power law of HI 21cm emission and absorption 
in our own galaxy, HI 21cm emission from the LMC and SMC 
are all $\sim -3$ (Crovisier \& Dickey (1983); Green (1993);
Deshpande et al. (2000); Elmegreen et al. (2001); Stanimirovic et al. (1999)).
Similarly, the HI distribution in several dwarf galaxies in the M81 group
appears to be fractal (Westpfahl et al.(1999)), consistent with
there being no preferred length scale. This scale free 
behaviour is characteristic of a turbulent medium; it is believed
that the turbulence itself is driven by a combination energy
input from spiral arms/bars and energy input from stellar
sources (supernovae, stellar winds etc.).

    Previous quantitative measurements of the HI emission power
spectrum have been limited to the few cases (the Milkyway, LMC
and SMC) where the HI signal is extremely strong. In this paper,
we discuss the power spectrum of the fluctuations of HI 21cm 
emission intensity in DDO~210, the faintest (M$_{\rm B} \sim -10.9$)
gas rich member of the local group. We use a visibility based 
estimator of the power spectrum; this is well suited to the
current problem where the signal is buried deep in the noise. 
Visibility based  methods also have the advantage of having 
well understood statistics, and being free of the uncertainties 
involved in gridding and image deconvolution. The rest of the 
paper  is divided as  follows. The power spectrum estimator 
that we use is discussed in detail in Sect.~\ref{sec:method}, 
the DDO~210 data presented in Sect.~\ref{sec:data},  while the 
results of applying the estimator to this data are shown 
in Sect.~\ref{sec:result} and discussed in 
Sect.~\ref{sec:discuss}.

\section{ A visibility based power spectrum estimator}
\label{sec:method}

The power spectrum of HI emission $\rm{P_{\rm{HI}}}(u,v)$ is the  
Fourier transform of the autocorrelation function 
$\xi(l-l^\prime,m-m^\prime)$
of  the fluctuations in the HI brightness  distribution 
$\delta I(l,m)$ {\rm i.e.}
\begin{equation}
\xi(l-l^\prime,m-m^\prime)=
\langle \delta I(l,m) \delta I(l^\prime, m^\prime) \rangle
\label{eq:a1}
\end{equation}
and 
\begin{equation}
{\rm{P_{\rm{HI}}}}(u,v)=\int \int~\xi(l,m)~
e^{- i 2\pi(ul+vm)}~dl~dm
\label{eq:a2}
\end{equation}
Here $(l,m)$ refers to  directions  on the sky and
  $(u,v)$ to inverse angular separations. The Fourier relation
in Eqn.~\ref{eq:a2} assumes that the angular extent of the  galaxy is 
small,   a spherical harmonic  decomposition would be needed otherwise. 
Further, it is assumed that the statistical properties of
the small scale fluctuations in the HI distribution are 
homogeneous and isotropic. The angular brackets denote
an average across different positions and directions in
the galaxy.  It follows that  the HI power 
spectrum ${\rm P}_{\rm HI}(\vec{U})$ is a function of only the 
magnitude $U=\sqrt{u^2+v^2}$. 

 Now, the complex visibility function measured by an interferometer 
is the Fourier  transform of the brightness distribution,
\begin{equation}
V(u,v)=\int \int I(l,m)~
e^{- i 2\pi(ul+vm)}~dl~dm \, .
\label{eq:a3}
\end{equation}
It should be noted that $(u,v)$ in Eqn.~\ref{eq:a3} 
refers to the projected baselines or antenna separations 
in units of the wavelength of observation. The 
Fourier relation in Eqn.~\ref{eq:a3} allows us to 
associate an inverse angular scale with every baseline $(u,v)$
which we also denote using a two dimensional vector $\vec U$.  
Since the complex visibility measured by an radio interferometer is the 
Fourier transform of the HI intensity distribution, the squared modulus
of the visibility is a direct estimator of the power spectrum,
\begin{equation} 
\langle V(\vec{U}) V^*(\vec{U})
 \rangle  =  {\rm P}_{\rm HI}(\vec{U})\,
\end{equation}
where  $\langle ... \rangle$ now denotes an average over 
all possible orientations  
of the baselines $\vec{U}$.  

 This estimator has been used by both Crovisier \& Dickey(1983) 
and Green(1993) to measure the power spectrum of HI fluctuations 
in our own galaxy. Eqn.~\ref{eq:a3} is an idealised one, in practice
the interferometer response differs from the Fourier transform of
the sky brightness both because it is modified by a window 
function and because there is added noise. We first consider 
the effect of the window function. In situations like in our 
own galaxy, where the HI  emission fills the whole telescope beam, 
the window function is simply the primary beam of the telescope. 
In situations where the HI emission is from, for e.g., a distant 
galaxy whose angular extent is much smaller than the primary beam,
the window function corresponds to the total extent of the 
HI emission. We will assume that, in general, the HI brightness 
distribution at any  frequency $\nu$ can be written as

\begin{equation}
I_\nu (l,m)=W_\nu(l,m)~[\bar{I}_{\nu}+ \delta I_{\nu}(l,m)]
,
\end{equation}
where $W_\nu$(l,m) is the window function which characterises the large 
scale HI distribution in the galaxy at frequency $\nu$, $\bar{I}_{\nu}$ 
is the mean HI  brightness and $\delta I_{\nu}(l,m)$ characterises the 
small scale fluctuations.  For a window function with angular extent 
$1/D$, we expect its Fourier transform, $\tilde W (\vec U)$, to 
be sharply peaked at 
$\vec U=0$ and to fall  off rapidly for $\mid \vec{U} \mid \gg D$.
Then, at long baselines, where we can ignore the contribution from
$W_\nu(l,m)~\bar{I}_{\nu}$, the square of the complex visibility  
is the power spectrum $P_{HI}(\vec U)$ convolved  with the square 
of the Fourier transform of the window function, i.e.
\begin{equation}
\langle V_\nu (\vec U)~V_\nu ^*(\vec U) \rangle=
{\tilde W_\nu}^2(\vec U)~\otimes~P_{HI}(\vec U) \,.
\label{eqn:first}
\end{equation}
Both Crovisier \& Dickey (1983) and Green(1993) ignored the window 
function in their analysis, which is justified at large baselines if 
${\tilde W_\nu}^2(\vec U)$   is sharply peaked at
 $\vec U=0$ and   decays much faster than  ${\rm P}_{\rm HI}(\vec{U})$. 

We now look at the effect of measurement noise. The measured visibility can 
be written as
$V_\nu (\vec U)= S_\nu (\vec U)+ 
N_\nu$, where $S_\nu (\vec U)$ is the HI signal and  
$N_\nu$ is the noise. We then have 

\begin{equation}
\langle V_\nu (\vec U)~V_\nu ^*(\vec U) \rangle=\langle S_\nu(\vec U) S^*_\nu(\vec U) \rangle+\langle N_\nu N^*_\nu \rangle
\label{eqn:second}
\end{equation}
where we see that  squaring the complex visibility makes the noise bias 
positive-definite. For HI emission from our own galaxy this noise bias 
is small compared to the signal. On the other hand,  the noise bias can 
be orders of magnitude larger than the desired power spectrum in the
case of faint external galaxies.  In principle, one could  estimate 
$\langle N_\nu N^*_\nu \rangle$ from a line free frequency  channel 
and subtract it from 
$\langle V_\nu (\vec U)~V_\nu ^*(\vec U) \rangle$ for  a
channel with emission.  However this requires that the noise statistics 
at different frequency channels be determined at an extremely high level 
of precision. Uncertainties in the bandpass response usually render 
such high levels of precision unachievable. 

The  problem of noise bias can be avoided by correlating the visibilities
at two different baselines for which  the noise is expected to be 
uncorrelated.  The correlation between the visibilities measured on two 
slightly different baselines is:

\begin{eqnarray*}
\hat{\rm P}_{\rm HI}(\vec U, \Delta \vec U)=
\langle V_\nu (\vec U)~V_\nu ^*(\vec U + \Delta \vec U) 
\rangle
\end{eqnarray*}

\begin{equation}
~~~~~~= \int {\rm P}_{\rm HI}(\vec{U}^{'}) [\tilde W_\nu(\vec U-\vec U^{'}) \times \tilde W^*_\nu(\vec U + \Delta \vec U-\vec U^{'})]~d^2 \vec U^{'},
\label{eqn:corr}
\end{equation}

Since $\tilde W (\vec U)$ falls  off rapidly for $\mid \vec{U} \mid \gg D$,
it then follows that the two shifted window functions in Eqn.~\ref{eqn:corr}
will have a substantial overlap only if $|\Delta \vec U| < D$. 
Thus visibilities at two  different baselines  will be correlated only
as  long as $|\Delta \vec U| < D$, and not beyond. In 
our analysis we restrict the difference in baselines to 
$|\Delta \vec U| \ll  D$ so that 
$\tilde W_\nu(\vec U + \Delta \vec U-\vec U^{'})
\approx \tilde W_\nu(\vec U-\vec U^{'})$. 
The expected value of the visibility correlation now no
 longer depends on $\Delta \vec U$ allowing us to introduce the notation
$\hat{\rm P}_{\rm HI}(\vec{U})=
\langle V_\nu (\vec U)~V_\nu ^*(\vec U + \Delta \vec U)$ and 

\begin{equation}
\hat{\rm P}_{\rm HI}(\vec{U})
= \int {\rm P}_{\rm HI}(\vec{U}^{'}) \mid\tilde W_\nu(\vec U-\vec U^{'})
\mid^2 ~d^2 \vec U^{'}, 
\label{eqn:corrn}
\end{equation}

We use the real part of the measured visibility 
correlation $\hat{\rm P}_{\rm HI}(\vec{U})$ as an estimator
 of the HI power 
spectrum  ${\rm P}_{\rm HI}(\vec{U})$. This is the  sum  of
the HI signal
and a system noise contribution. The latter 
is unbiased and has a zero mean, its 
variance being  determined by the noise statistics 
of the individual visibilities. The imaginary part of 
$\hat{\rm P}_{\rm HI}(\vec{U})$ is noise
 dominated. There will be a very small 
contribution from the HI signal because the
assumption that 
$\tilde W_\nu(\vec U + \Delta \vec U-\vec U^{'})
\approx \tilde W_\nu(\vec U-\vec U^{'})$
is not strictly valid.  The requirement that the 
imaginary part of  $\hat{\rm P}_{\rm HI}(\vec{U})$ should be small
provides a self-consistency  check and allows us to determine the
 range  of validity of our formalism.

A further simplification is possible for our estimator  if we assume
that  
$\tilde{W}_\nu(\vec U)$  decays much faster  than variations in ${\rm
  P_{\rm HI}}(\vec{U}^{'})$.  
Such an assumption is
justified at large baselines 
$U \gg D$ if, for example, ${\rm P}_{\rm HI}(\vec{U})$ 
is a power law. We then have  

\begin{equation}
\hat{\rm P}_{\rm HI}(\vec{U})=
 {\rm P}_{\rm HI} (\vec{U}) \, \left[
\int  \mid \tilde W_\nu(\vec U^{'}) \mid^2 
~d^2 \vec U^{'}\right] \,.
\label{eqn:corra}
\end{equation}
where the quantity in square brackets $[..]$ in Eqn.~\ref{eqn:corra} is a constant.  
We note that while $\hat{\rm P}_{\rm
  HI}(\vec{U}) $ directly estimates $P_{\rm HI}$ at large baselines $U \gg
D$, it is necessary to account for the convolution
(Eqn. \ref{eqn:corrn}) at small baselines $U \sim D$.

    The utility of using visibility correlations as a statistical
estimator of the HI power spectrum has been discussed  previously 
by Bharadwaj \& Sethi (2001) in the context of the large scale HI 
distribution at high redshifts. Visibility based statistical 
techniques are also of interest in the efforts to detect the 
epoch of reionization through interferometric HI observations
(eg. \citealt{morl};\citealt{BA5}), and are also used in the 
analysis of interferometric observations of the CMBR  
(e.g. \cite{hobson95}).

\section{Data}
\label{sec:data}

\begin{figure*}
\psfig{file=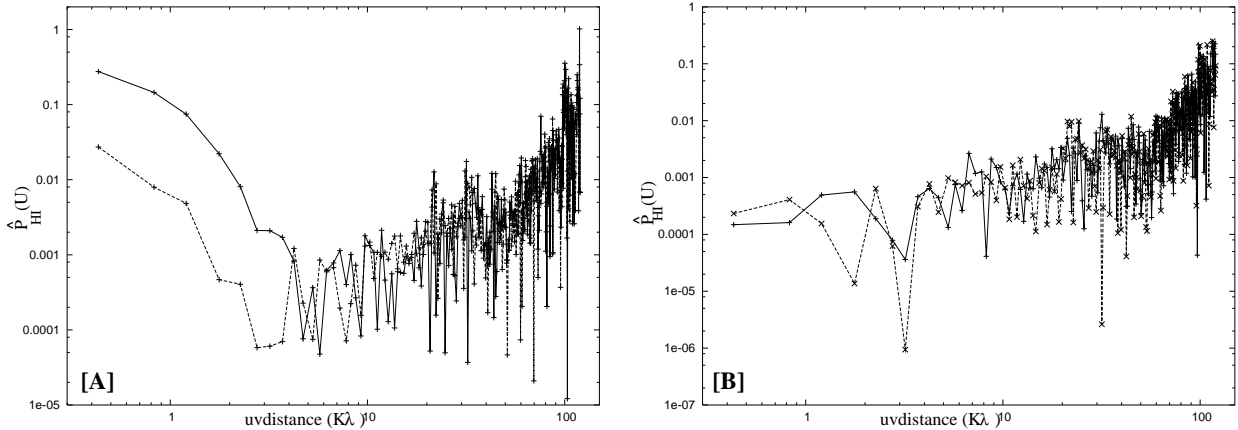,width=6.6truein}
\caption{
Absolute value of the real (solid line) and imaginary (dash line) parts
 of $\hat{\rm P}_{\rm HI}(\vec{U})$  plotted  against the uv
distance $U$ for  two
different frequency channels of the GMRT data. {\bf [A]} shows
results for a channel with significant HI emission and {\bf [B]}  a
channel with no HI emission,   
corresponding  to  heliocentric velocities  of
$\sim -$137 kms$^{-1}$ and  $\sim -64$  kms$^{-1}$ respectively.}
\label{fig:gmrt_ps}
\end{figure*}

     We apply the formalism developed above to interferometric observations
of DDO~210. Since DDO~210 is relatively close (at a distance of 
950$\pm$50 kpc; \cite{lee99}) the observations have good spatial 
resolution as well as 
comparatively good  signal-to-noise ratio (SNR). Further, DDO 210 has 
been observed with both the Giant Metrewave Radio Telescope (GMRT) (Swarup et al. 1991) 
and in multiple configurations of the Very Large Array (VLA). A comparison
of the power spectrum derived from these two independent data sets allows
a robust check against systematic biases. 

     The GMRT data for DDO 210 have been discussed in detail in 
\cite{begum04}. We state the main observational results briefly
below, the reader is referred to \cite{begum04} for a more details. 
The HI emission  from DDO 210  spanned the central 25 channels 
of the 128 channel  spectral cube (with channel width $\sim$1.7 kms$^{-1}$). 
The HI disk of the galaxy is  nearly face on. On large scales, 
the  HI distribution is not axisymmetric; the integrated HI 
column-density contours are elongated towards the east and south.
The  maximum angular extent of the HI distribution for the galaxy 
is $D \sim 0.8~\rm{k}\lambda$ and does not change significantly 
in the central 10 frequency channels.  

     The VLA observations have been discussed in \cite{young03}.
For the current analysis  we use raw data downloaded from the 
VLA archive and  reanalysed by us.

\section{Results}
\label{sec:result}

   We estimated $\hat{\rm P}_{\rm HI}(\vec{U}) $ by correlating all the 
baselines within $\mid~\Delta \vec U~\mid \le 0.2 \, \rm{k} \lambda$. 
This  value of $\Delta \vec U$ was chosen as a  compromise between
having sufficient visibility pairs to get a good SNR and ensuring  
that $\Delta \vec U$  itself remains small compared to  $D\sim0.8~\rm{k} \lambda$. 
All self correlations (i.e. correlations of a visibility with itself)
were excluded. The  measured correlations were then binned in bins
of $0.5~\rm{k}\lambda$ width. Each frequency channel was treated separately,
and  the analysis was also carried out on the line free frequency 
channels as a check. 

    Fig.~\ref{fig:gmrt_ps}[A] shows the absolute value of the real 
and imaginary parts of  $\hat{\rm P}_{\rm HI}(\vec{U}) $  for a channel 
with HI emission, while Fig.~\ref{fig:gmrt_ps}[B] shows the same 
quantities for a channel with no emission. For the channel with
emission, we find that at baselines less than $\sim 5~\rm{k}\lambda$ 
the real and imaginary parts of  $\hat{\rm P}_{\rm HI}(\vec{U})$
are both positive, and the real part is substantially larger 
than the imaginary part. For the channel with no emission the
real and imaginary parts are small and of comparable 
magnitude. This is in agreement with the cross-checks 
discussed in Sec.~\ref{sec:method} for determining the reality 
of the signal. Henceforth, we will ignore the imaginary part
of $\hat{\rm P}_{\rm HI}(\vec{U}) $ and use the symbol 
$\hat{\rm   P}_{\rm HI}(\vec{U}) $ to refer to only the real
part. Note that the apparent increase in the value  
of $\hat{\rm P}_{\rm HI}(\vec{U}) $ seen at the longer baselines in
Fig.~\ref{fig:gmrt_ps}[A]\&[B]  is because the noise is large
in this range and we are plotting the absolute value of the
estimator. The increase in noise is a consequence of the
sparse sampling of the outer parts of the uv-plane. 

\begin{figure}
\psfig{file=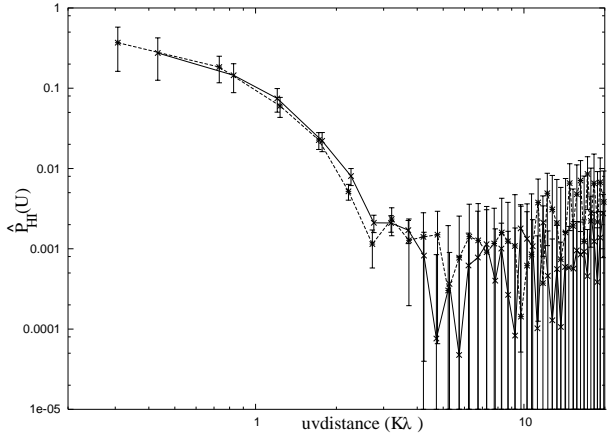,width=3.3truein}
\caption{Absolute value of the real part  of  $\hat{\rm P}_{\rm HI}(\vec{U})$     
plotted as a function of uv-distance $U$ for the VLA data (dash line)
and the GMRT data (solid line) for a frequency channel with significant HI emission. 
The GMRT data is exactly the same as in  Fig.~\ref{fig:gmrt_ps}, also  the VLA
frequency channel has  the same  heliocentric velocities as  the GMRT data.}
\label{fig:vla_ps}
\end{figure}

Fig.~\ref{fig:vla_ps} shows $\hat{\rm P}_{\rm HI}(\vec{U}) $ determined
separately from the GMRT and VLA data (for channels corresponding to 
very similar heliocentric velocity). The two estimates are in excellent 
agreement, which is a robust check on the reality of the signal. 
Fig.~\ref{fig:vla_ps} also shows  $1\sigma$ error bars for 
the estimated power spectrum. The errors were computed by including both
the effect of having a fine number of estimates of the true spectrum
(``cosmic variance'') as well as having noisy visibility measurements.
The first term was estimated as  $\hat{\rm P}_{\rm HI}(\vec{U}) \sqrt{D/2\pi U}$,
while the second term was taken from the measured fluctuations of 
$\hat{\rm P}_{\rm HI}(\vec{U}) $ for channels with no HI emission. 
The total $1\sigma$ error bars were determined by  combining the 
two contributions in quadrature.  At small baselines $(U \le 5\, \rm{k} \lambda)$,
the first term dominates the error budget, while the noise 
on the measured visibilities dominates at large baselines.

\begin{figure} 
\psfig{file=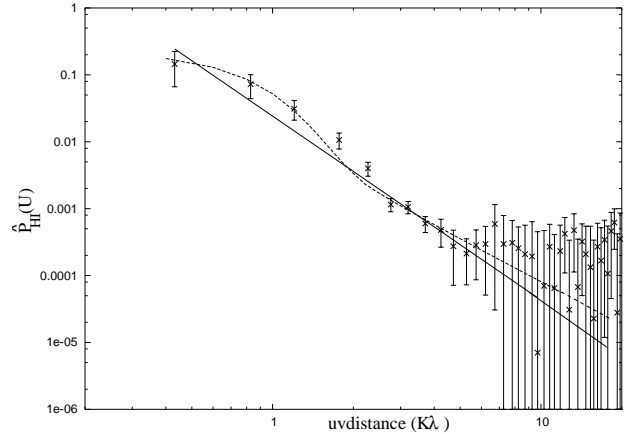,width=3.3truein}
\caption{ Absolute value of the real part  of  $\hat{\rm
  P}_{\rm HI}(\vec{U}) $ 
 determined after averaging the GMRT visibilities over
the 10 central frequency channels which contain significant HI
emission.  The best fit power law using the UV range $2 \, -\, 12 \,
\kl$ is shown as a solid line. The best fit convolved power law
over the UV range $0 \, -\, 12 \, \kl$ is shown as a dotted line.
See the text for more details. }
\label{fig:powerlaw_ps}
\end{figure}

    If the observed power spectrum is due to turbulence in the  
interstellar medium, the slope of the derived power spectrum  
could change depending on the  width of the frequency channel
(\cite{lazarian95}). Turbulence gives rise to fluctuations in
both the density and  velocity of the gas and both of these
contribute  to the observed  intensity fluctuations. Lazarian \&
Pogosyan (2000) show that the statistical properties of
the velocity and the density fluctuations can be disentangled  if 
the slope of the observed power spectrum changes depending on
whether the intensity is averaged over a velocity range that
is large (``thick slices'') or small (``thin slices'') compared
to the turbulent velocity dispersion. For DDO~210 the observed 
velocity dispersion is $\sim 6.5$ \kms and is approximately 
constant across the galaxy.  $\hat{\rm P}_{\rm HI}(\vec{U}) $ 
estimated from visibilities averaged across the 10 channels 
over which the emission has a similar spatial distribution 
is shown in Fig.~\ref{fig:powerlaw_ps}. Within our measurement 
errors, we find no change in the power spectrum with increasing 
velocity width in the range from $\sim 1.7$ \kms to 
$\sim 17$ \kms. We discuss the implications of this in
Sec~\ref{sec:discuss}, but we note here that in this means that
one can work  with the emission averaged over channels
to improve the signal to noise ratio.

The power spectrum  $\hat{\rm P}_{\rm HI}(\vec{U}) $ shown in
Fig.~\ref{fig:powerlaw_ps} appears to fall like 
a power law beyond a uv distance of $\sim 1~\rm{k}\lambda$. 
The best fit power  law was determined by minimising 
\begin{equation}
\chi^2=\sum_i \frac{[\hat{\rm P}_{\rm HI}(U_i)-A
  U_i^{\alpha}]^2}{\sigma_i^2}
\label{eqn:chisquare}
\end{equation}
with respect to the amplitude $A$ and the power law index
$\alpha$. It is also necessary to fix the uv range while doing
the fit. As can be seen in Fig.~\ref{fig:powerlaw_ps}, 
$\hat{\rm P}_{\rm HI}(\vec{U})$ flattens  below $1~{\rm k}\lambda$;
at these scales one would expect that the convolution with the
window function $|\tilde W_\nu(\vec U)|^2$ would be important.
If we assume that the convolution with the window function
can be ignored for $U > 1~ \rm{k}\lambda$ and do the minimisation
described in Eqn.~\ref{eqn:chisquare} with the lower limit fixed
at $U_{\rm min} = 1~\rm{k}\lambda$,  but several different upper limits
$U_{\rm  max}$,  we find that the minimum $\chi^2$/DOF 
decreases up to $U_{\rm max}=12~{\rm k}\lambda$ but not beyond.  For the 
range $1-12$ k$\lambda$ the best fit power law index is 
$\alpha$=$-3.55^{+0.40}_{-0.30}$. Similarly, keeping  $U_{\rm min} = 
2~\rm{k}\lambda$, the best fit power law index for the uv-range 
$2-12$ k$\lambda$ is $\alpha$=$-2.85^{+0.50}_{-0.65}$.
The best fit for the uv-range $2-12$ k$\lambda$ is shown in 
Fig.~\ref{fig:powerlaw_ps}. We note that the  best fit
$\alpha$ is largely determined  by the smallest baselines $U \approx
U_{\rm min}$ as these have the smallest error bars, and it 
does not change significantly if $U_{\rm max}$ is changed to say $6 \,
{\rm k}\lambda$ or  even $50 \,{\rm k} \lambda$. As one goes to
larger and larger baselines however, the minimum in the $\chi^2$
distribution gets more and more shallow -- equivalently the
error bars on $\alpha$ get larger and larger.

  To check the assumption that the convolution with the window
function can be ignored for $U>1~\rm{k}\lambda$, we modelled the 
window function  using a Gaussian $\tilde W_\nu(\vec U)=e^{-U^2/2 D^2}$ 
of   width  $D=0.8~{\rm k}\lambda$.  For a power law form 
of ${\rm P}_{\rm HI}(\vec{U})$ we find that the convolution actually
produces measurable  deviations from  the power law  at 
$U \le 2~{\rm k}\lambda$. The observed power spectrum is 
{\it steeper} than the underlying power law spectrum in the 
range $1-2~{\rm k}\lambda$, while at baselines $U \le 1~{\rm k}\lambda$
it is {\it flatter} than the underlying power law.  We hence repeated
the least squares fit to the observed power spectrum using a power 
law convolved with the assumed Gaussian window function as the
input model. For this model the best fit slope for the range
$U \le 12 \kl$  is $\alpha=-2.1 ^{+0.05}_{-0.10}$. The value of 
the best fit  slope when including baselines $<2 \kl$ could depend
on  how well  the true window function has been modelled. In
our analysis we have approximated this with a Gaussian;
any deviations from a Gaussian could introduce unknown biases in our
estimate of the  slope. With this consideration in mind, we also
did fits for the restricted uv-range $2-12 \kl$. For this uv-range, 
the best fit slope is $\alpha=-2.75 \pm0.45$. While the fit is good
over the fitting range, the model substantially over estimates the 
power on large scales. This is consistent with the flatter slope
($\sim -2.1$) obtained when fitting over the entire $U \le 12 \kl$
range, and might be indicative of a deviation from a power law 
on large scales. However, given the uncertainty in the modelling of 
the window function it is difficult to make
a quantitative statement. On the other hand, our modelling does
indicate that beyond about $\sim 2 \kl$ the effect
of the window function is marginal. The value of the
slope obtained after taking into account  the convolution 
(viz. $-2.75 \pm0.45$) is consistent within the error bars with
the value $\alpha= -2.85^{+0.50}_{-0.65}$ obtained without 
accounting for the convolution. We hence adopt the value 
$\alpha=-2.75 \pm0.45$ as slope of the intrinsic power law
over these spatial scales.

\section{Discussion}
\label{sec:discuss}

        Over the uv-range $2 \, -\, 12 \, \kl$ (corresponding to spatial
scales of 80pc - 500pc) the power spectrum of  HI intensity fluctuations 
in the dwarf galaxy DDO 210 is well described by a  power law with
slope $\alpha=-2.75 \pm0.45$.  This value is very close to the 
slope of $\sim- 3$ estimated from the  HI emission in our own galaxy  
(on spatial scales of $\sim$ 5 pc to 200 pc) by Crovisier \&
Dickey(1983) and Green(1993), and for the SMC (on spatial scales
between $\sim$30~pc to 4~kpc; Stanimirovic et al. 1999). The reason why
the HI emission power spectrum has a similar slope in the Milkyway
as in a galaxy with more than 3 orders of magnitude less HI mass 
suggests that some fundamental physical process is responsible
for determining the power spectrum slope. Power spectral slopes $\sim -3$ 
are expected from optically thick turbulent gas (\cite{lazarian04}).
The HI emission from late type dwarfs 
like DDO~210 is however expected to be optically thin, (see e.g. 
\cite{haynes84} \citealp{giovanelli94}).  Models like those
of \cite{goldman00} which give a spectral slope $\sim -3$
in optically thin gas may be more relevant in this case.

 Unlike the Milkyway, DDO~210 has no spiral arms and it also 
has no measurable star formation (\cite{vanzee00}).
   The lack of dependence of the power spectrum slope on the star 
formation rate that we find is 
similar to the finding of Dib et al. (2006) that the characteristic 
HI velocity dispersion in galaxies shows a minimum level that is 
independent of the star formation rate. In this aspect,
the turbulence in the gas distribution appears to be 
different from that in the stellar distribution, where 
the power spectrum's slope does seem to depend on the star 
formation rate (Willett 2005). It is possible that the
energy source driving the HI turbulence is gravitational
energy input on large scales.  Detailed simulations suggest
that large scale gravitational energy input could explain
the substantial turbulent velocity dispersion seen in the
outer parts of the extended gas disk of NGC~2915 (Wada et al. (2002))
as well as for the rich structures  seen in Holmberg~II 
(Dib \& Burket (2005)). Gravitational energy input
on large scales (perhaps by tidal torques from other
local group members) would also be consistent with the
asymmetric large scale HI distribution in DDO~210.

   As mentioned in Sec.~\ref{sec:result}, turbulence produces 
fluctuations in both the density and velocity fields, in turn,
both of these contribute to fluctuations of the HI intensity.
Lazarian \&Pogosyan (2000) show that the slope of the HI
intensity power spectrum  goes like $n+\zeta_2/2$ when measured
for  channel widths that are ``thin''  compared to the 
turbulent velocity dispersion, while for channel widths
that are ``thick'' compared to the velocity dispersion
the HI intensity power spectrum's slope is $n$, where $n$ 
is the slope of the power spectrum of density fluctuations 
and $\zeta_2$ is the spectral slope of the velocity structure function.
Essentially, for velocity widths large 
compared to the velocity dispersion (``thick'' slices)
the  velocity information gets averaged out and the 
intensity fluctuations represents density fluctuations. 
Recall that we find that within the measurement errors the
slope of the HI power  spectrum  does not change for 
velocity widths ranging from $\sim 1.7$~\kms to $\sim 17$~\kms. 
The observed HI velocity dispersion of $6.5$ \kms has 
contributions  from both thermal and turbulent random 
motions. Our  ``thick'' channel is  wider than the total 
velocity dispersion, and hence we can be sure that
the detected  power spectrum represents the power spectrum
of the HI density fluctuations. Since we get the same 
slope $\sim -2.75 \pm 0.45$  even for $\sim 1.7$ \kms 
channels, either the  turbulence velocity dispersion 
is  smaller  than $1.7$ \kms or $ \zeta_2 \approx 0  \pm 0.9$.
Note that the latter possibility is consistent with $\zeta_2=2/3$ 
which is predicted by the Kolmogorov theory of turbulence. 

{\it Acknowledgments}
    The data presented in this paper were obtained using the GMRT 
(operated by the National Centre for Radio Astrophysics of the 
Tata Institute of Fundamental Research) and the NRAO VLA.
The National Radio Astronomy Observatory is a facility of 
the US National Science Foundation operated under cooperative 
agreement by Associated Universities, Inc. SB would also 
like to acknowledge BRNS, DAE, Govt. of India for 
financial support through sanction No. 2002/37/25/BRNS.


\begin{thebibliography}{}
\bibitem[\protect\citeauthoryear{Begum \& Chengalur}{2004}]{begum04}
        Begum, A., \& Chengalur, J. N., 2004, A\&A, 413, 525
\bibitem[\protect\citeauthoryear{Bharadwaj \& Sethi}{2001}]{bharadwaj02} Bharadwaj, S. \& Sethi, S. K., 
        2001, JA\&A, 22, 293
\bibitem [\protect\citeauthoryear{Bharadwaj and Ali }{2005}]{BA5}
  Bharadwaj S. \& Ali S. S., 2005, \mnras, 356, 1519
\bibitem[\protect\citeauthoryear{Crovisier \& Dickey}{1983}]{crovisier83} Crovisier, J. \& Dickey, J. M., 
        1983, A\&A, 122, 282
\bibitem[Deshpande et al.(2000)]{2000ApJ...543..227D} Deshpande, A.~A., 
Dwarakanath, K.~S., \& Goss, W.~M., 2000, ApJ, 543, 227 
\bibitem[Dib et al. (2006)]{dib06} Dib S., Bell, E. \& Burkert A.,
   2006, ApJ, 638, 797.
\bibitem[Dib \& Burkert (2005)]{dib05} Dib S. \& Burkert A., 2005, ApJ, 630, 238
\bibitem[\protect\citeauthoryear{Elmegreen et al.}{2001}]{elmegreen01} Elmegreen, B. G., Kim, S. \& 
        Staveley-Smith, L., 2001, AJ, 548, 749
\bibitem[\protect\citeauthoryear{Elmegreen \& Scalo}{2004}]{elmegreen04} Elmegreen B. G. \& Scalo J., 2004, ARAA, 42, 211.
\bibitem[\protect\citeauthoryear{Green}{1993}]{green93} Green, D. A., 1993, MNRAS, 262, 327
\bibitem[\protect\citeauthoryear{Hobson et al.}{1995}]{hobson95} Hobson, M. P., Lasenby, A. N. 
        \& Jones, M., 1995, MNRAS, 275, 863
\bibitem[\protect\citeauthoryear{Lazarian \& 
Pogosyan}{2004}]{lazarian04} Lazarian A., Pogosyan D., 2004, ApJ, 
616, 943 
\bibitem[\protect\citeauthoryear{Lazarian \& Pogosyan}{2000}]{lazarian00} Lazarian, A. \& Pogosyan, D., 
        2000, ApJ, 537, 720
\bibitem[\protect\citeauthoryear{Lazarian}{1995}]{lazarian95} Lazarian, A., 1995, A\&A, 293, 507
\bibitem[\protect\citeauthoryear{Giovanelli et 
al.}{1994}]{giovanelli94} Giovanelli R., Haynes M.~P., Salzer J.~J., 
Wegner G., da Costa L.~N., Freudling W., 1994, AJ, 107, 2036 
\bibitem[\protect\citeauthoryear{Goldman}{2000}]{goldman00} 
Goldman I., 2000, ApJ, 541, 701 
\bibitem[\protect\citeauthoryear{Haynes \& 
Giovanelli}{1984}]{haynes84} Haynes M.~P., Giovanelli R., 1984, 
AJ, 89, 758 
\bibitem[\protect\citeauthoryear{Lee et al.}{1999}]{lee99} Lee, M. G., Aparicio, A., Tikonov, N., Byun, Y.
        \& Kim, E., 1999, AJ, 118, 853.

\bibitem [\protect\citeauthoryear{Morales and Hewitt}{2004}]{morl}
  Morales,M. F. and Hewitt,J., 2004, ApJ, 615, 7
\bibitem[\protect\citeauthoryear{Scalo \& Elmegreen }{2004}]{scalo04} Scalo J. \& Elmegreen B. G., 2004, ARAA, 42, 275. 
\bibitem[\protect\citeauthoryear{Stanimirovic et al.}{1999}]{stanimirovic99} Stanimirovic, S., 
        Staveley-Smith, L., Dickey, J. M., Sault, R. J., \& Snowden, S. L., 1999, MNRAS, 302, 417
\bibitem[\protect\citeauthoryear{Swarup et al.}{1991}]{swarup91} Swarup, G., Ananthakrishnan, S.,
        Kapahi, V. K., Rao, A. P., Subrahmanya, C. R., \& Kulkarni, V. K., 1991, Current Science, 
	60, 95
\bibitem[van Zee (2000)]{vanzee00} van Zee L., 2000, AJ, 19 2757 
\bibitem[Wada et al. (2002)]{wada02} Wada K., Meurer, G. \& Norman C. A., 2002,
   ApJ, 577, 197
\bibitem[\protect\citeauthoryear{Westpfahl et al.}{1999}]{westpfahl99} Westpfahl, D. J., Coleman, 
         P. H., Jordan, A. \& Thomas, T., 1999, AJ, 117, 868
\bibitem[\protect\citeauthoryear{Willett et al.}{2005}]{willett05} Willett, K. W., Elmegreen, B. G. 
         \& Hunter, D. A., 2005, AJ, 129, 2186
\bibitem[\protect\citeauthoryear{Young et al.}{2003}]{young03}
        Young, L. M., van Zee, L., Lo, K. Y., Dohm-Palmer, R. C. \&
        Beierle, M. E., 2003, ApJ, 592, 111
\end{thebibliography}
\end{document}